\documentclass[%
aps,
pre,
 amsmath,amssymb,
reprint,%
]{revtex4-1}
\usepackage{graphicx}
\usepackage{dcolumn}
\usepackage{bm}

\usepackage{amsmath}	
\begin{document}

\preprint{AIP/123-QED}

\title{Vibrational density of states of jammed
packing at high dimensions: mean-field theory}
\author{Harukuni Ikeda}
 \email{harukuni.ikeda@gakushuin.ac.jp}
\affiliation{ 
Department of Physics, Gakushuin University, 1-5-1 Mejiro, Toshima-ku, Tokyo 171-8588, Japan}
\author{Masanari Shimada}
\affiliation{ 
Department of Physics, Toronto Metropolitan University, M5B 2K3, Toronto, Canada}

\date{\today}
 
\begin{abstract}
Several mean-field theories predict that the Hessian matrix of amorphous
solids converges the Wishart matrix in the limit of the large spatial
dimensions $d\to\infty$. Motivated by these results, we here calculate
the density of states of random packing of harmonic spheres by mapping
the Hessian of the original system to the Wishart matrix. We compare
our result with that of previous numerical simulations of harmonic
spheres in several spatial dimensions $d=3$, $5$, and $9$. For small
pressure $p\ll 1$ (near jamming), we find a good agreement even in
$d=3$, and obtain better agreements in larger $d$, suggesting that the
approximation becomes exact in the limit $d\to\infty$. 
\end{abstract}

\maketitle

\newcommand{\diff}[2]{\frac{d#1}{d#2}}
\newcommand{\pdiff}[2]{\frac{\partial #1}{\partial #2}}
\newcommand{\fdiff}[2]{\frac{\delta #1}{\delta #2}}
\newcommand{\bx}{\bm{x}}
\newcommand{\bq}{\bm{q}}
\newcommand{\br}{\bm{r}}
\newcommand{\ba}{\bm{a}}
\newcommand{\by}{\bm{y}}
\newcommand{\bY}{\bm{Y}}
\newcommand{\bF}{\bm{F}}
\newcommand{\bn}{\bm{n}}
\newcommand{\be}{\bm{e}}
\newcommand{\new}{\nonumber\\}
\newcommand{\abs}[1]{\left|#1\right|}
\newcommand{\tr}{{\rm Tr}}
\newcommand{\HH}{{\mathcal H}}
\newcommand{\II}{{\mathcal I}}
\newcommand{\WW}{{\mathcal W}}
\newcommand{\OO}{{\mathcal O}}
\newcommand{\ave}[1]{\left\langle #1 \right\rangle}
\newcommand{\im}{{\rm Im}}
\newcommand{\re}{{\rm Re}}
\newcommand{\ke}{k_{\rm eff}}

\section{Introduction}

The vibrational density of states $D(\omega)$ plays a central role to
characterize the low-temperature properties of solids. For both crystals
and amorphous solids, $D(\omega)$ for small $\omega$ eventually follows
the prediction of the Debye model $D(\omega)\sim \omega^{d-1}$,
suggesting that the vibrational excitation is dominated by phonon
modes~\cite{kittel1976introduction,mizuno2017continuum}. However, for
amorphous solids, in addition to the phonon modes, there arise excess
non-phonon modes for small $\omega$. This phenomenon, often referred to
as the boson peak, is considered as a universal feature of amorphous
solids~\cite{phillips1981amorphous}.

From the theoretical point of view, a first step to tackle the problem
is to consider mean-field models/theories. Several mean-field models,
such as the $p$-spin spherical model~\cite{biroli2012random} and
perceptron~\cite{franz2015universal}, and theories, such as the
effective medium theory~\cite{degiuli2014effects,shimada2020random},
cavity method~\cite{parisi2014soft},
etc.~\cite{beltukov2015random,cicuta2018,baggioli2019,PhysRevResearch.1.012010},
suggest that Hessian matrices of amorphous solids are approximated by
the Wishart matrix. However, somewhat surprisingly, the functional form of
$D(\omega)$ of particle systems has not been calculated yet, even in the
large dimensional limit $d\to\infty$, where the mean-field theory
becomes exact. As a consequence, one should introduce fitting parameters
to compare the theory and numerical
results~\cite{manning2015random,beltukov2015random,baggioli2019}, even
in large $d$~\cite{PhysRevLett.117.045503}.

In this work, we focus on frictionless spherical particles interacting
with the harmonic potential~\cite{ohern2003}. Since the harmonic
potential is a purely repulsive potential, the system gets unstable in
the zero pressure limit $p\to 0$, which is known as the (un)jamming
transition~\cite{ohern2003}. Near the jamming transition point ($p\ll
1$), several physical quantities, such as the contact number $z$,
exhibit the power-law behavior~\cite{ohern2003}. The critical exponents
near the jamming transition are calculated by several mean-field
theories~\cite{wyart2005,degiuli2014effects,franz2017universality,parisi2020theory}.
However, again, the detailed functional form of $z$ is still
undetermined, even in $d\to\infty$.

Recently, one of the present authors performed an extensive numerical
simulation of harmonic spheres and calculated $D(\omega)$ and $z$ in
spatial dimensions from $d=3$ to $d=9$~\cite{shimada2020low}. Therefore,
it is now desirable to directly compare the numerical results in large
$d$ with the predictions of the mean-field theory.

Here, we theoretically calculate $D(\omega)$ and $z$ of harmonic spheres
in large $d$, and compare them with the previous numerical results. For
this purpose, inspired by the previous mean-field calculations, we
assume that the Hessian of harmonic spheres converges to the (shifted)
Wishart matrix in the mean-field limit $d\to\infty$. We determine the
pre-factors of the Wishart matrix so that its trace is consistent with
that of the Hessian of the original model. For small pressure, our
results well agree with the previous numerical
results~\cite{shimada2020low} even in $d=3$, and obtain better
agreements in larger $d$, suggesting that our theory becomes exact in
the limit of $d\to\infty$.


The organization of the paper is as follows.  In
Sec.~\ref{170619_26Oct20}, we introduce the model and several physical
quantities.  In Sec.~\ref{170705_26Oct20}, we calculate $D(\omega)$ in
the limit of $d\to\infty$.
In Sec.~\ref{170931_26Oct20}, we summarize the results.

\section{Settings}
\label{170619_26Oct20} Here we introduce the model and several physical
quantities. We consider a system consisting of frictionless spherical
particles interacting with the harmonic potential~\cite{ohern2003}:
\begin{align}
& V = \sum_{i<j}^{1,N}k\frac{h_{ij}^2}{2}\theta(-h_{ij}),
 \ h_{ij} = \abs{\br_i-\br_j}-R_i-R_j
\end{align}
where $N$ denotes the number of particles, $k$ denotes the spring
constant, and $\theta(x)$ denotes the Heaviside step function. $\br_i =
\{x_{i1},\dots, x_{id}\}$ and $R_i$ denote the position and radius of
the $i$-th particle, respectively. To simplify the notation, hereafter,
we set $k=1$.

The Hessian of the potential is
\begin{align}
&\mathcal{H}_{ia,jb} =\pdiff{^2 V}{x_{ia}\partial x_{jb}}
  = \HH_{ia,jb}^{(1)} + \HH_{ia,jb}^{(2)},\new
&\HH_{ia,jb}^{(1)} = \sum_{\mu=1}^{Nz/2}
 \pdiff{h_{\mu}}{x_{ia}}\pdiff{h_{\mu}}{x_{jb}},\
\HH_{ia,jb}^{(2)} = \sum_{\mu=1}^{Nz/2}h_{\mu}
\pdiff{^2 h_{\mu}}{x_{ia}\partial x_{jb}},
\label{145125_2Sep20}
\end{align}
where
\begin{align}
 z = \frac{1}{N}\sum_{i<j}\theta(-h_{ij}).
\end{align}
denotes the number of contacts per particle, and $\sum_{\mu=1}^{Nz/2}$
denotes the sum of all pairs $\mu=(ij)$ for which $h_{ij}<0$.  Once we
have the eigenvalue distribution of $\HH$, $\rho(\lambda)$, the
vibrational density of states $D(\omega)$ is calculated as
\begin{align}
 D(\omega) = 2\omega\rho(\lambda=\omega^2).\label{134253_15Sep20}
\end{align}
For the control parameter, we use the \textit{pre-stress} defined
as~\cite{shimada2020low}
\begin{align}
 e = -\frac{2(d-1)}{Nz}\sum_{\mu=1}^{Nz/2}\frac{h_\mu}{r_\mu}
 = (d-1)\ave{\frac{R_i+R_j}{r_{ij}}-1},\label{144626_10May21}
\end{align}
where $\ave{\bullet}$ denotes the average for the all contacts
$\ave{\bullet} = (Nz/2)^{-1}\sum_{i<j}^{Nz/2}\theta(-h_{ij})\bullet$.
The right most expression in Eq.~(\ref{144626_10May21}) clearly shows
that $e$ proportional to the average overlap of particles. The
proportional constant $(d-1)$ has been chosen so that $e$ remains finite
in the limit $d\to\infty$~\cite{shimada2020low}. Near the jamming
transition point, $e$ is proportional to the pressure, $e\sim p$ and
vanishes at the jamming transition point. In a previous numerical
study~\cite{PhysRevLett.117.045503}, the packing fraction was used as a
control parameter. However, it has been pointed out that $e$ is a more
natural control
parameter~\cite{bi2015statistical,shimada2020low}. Below, we calculate
$z$ and $D(\omega)$ as functions of $e$.

\section{Theory}
\label{170705_26Oct20}

\subsection{Summary of previous works}

Here we briefly review the previous works. The seminal work has been
done by G.~Parisi~\cite{parisi2014soft}. He showed that the eigenvalue
distribution of the Hessian of harmonic spheres converges to the
Marcenko-Pastur distribution in the limit $d\to\infty$, meaning that the
Hessian is identified with the Wishart matrix $\HH \sim
\WW$~\cite{parisi2014soft}, but the effects of the pre-stress $e$ were
neglected at that time. More recently, G.~Parisi with coauthors
performed a more complete analysis for the perceptron, which is a
mean-field model of random sphere packing of harmonic
spheres~\cite{franz2016simplest,franz2017universality}. The analysis of
the perceptron suggests that the Hessian of harmonic spheres in large
$d$ is written as
\begin{align}
\HH_{\rm MF} = a \WW + b e\II,\label{131641_25Oct20}
\end{align}
where $a$ and $b$ denote constants, $\II_{ia,jb}=\delta_{ia,jb}$ denotes
the identity matrix, and
\begin{align}
 \WW_{ia,jb}=\frac{2}{Nz}\sum_{\mu=1}^{Nz/2}\xi_{ia}^\mu\xi_{jb}^\mu
\end{align}
denotes the Wishart matrix. $\xi_{ia}^\mu$ denotes the i.i.d gaussisan
random variable with zero mean and unit
variance~\cite{franz2015universal}. The replica calculation of the
perceptron also proves the marginal stability~\cite{muller2015marginal}:
the minimal eigenvalue $\lambda_{\rm min}$ of the Hessian $\HH$ vanishes
$\lambda_{\rm min}=0$ near the jamming transition
point~\cite{franz2015universal}. Interestingly, several other mean-field
theories also suggest that the Hessian of harmonic spheres is written as
Eq.~(\ref{131641_25Oct20}), though the precise values of $a$ and $b$ are
still
unknown~\cite{degiuli2014effects,beltukov2015random,cicuta2018,baggioli2019,ikeda2020}.

In this work, we use Eq.~(\ref{131641_25Oct20}) as an Ansatz.  We
determine $a$ and $b$ by requiring that the trace of $\HH_{\rm MF}$ is
consistent with that of the Hessian of the original model,
Eq.~(\ref{145125_2Sep20}).

\subsection{Calculation of $a$ and $b$}

For simplicity, we first discuss the case without pre-stress. By
setting $e=0$, we get
\begin{align}
 \HH \to \HH^{(1)},\ \HH_{\rm MF} \to a \WW.
\end{align}
Then, we determine $a$ from the following condition:
\begin{align}
 \tr \HH^{(1)} = a \tr \WW.\label{113356_13Feb21}
\end{align}
The LHS in Eq.~(\ref{113356_13Feb21}) can be calculated as 
\begin{align}
 \tr \HH^{(1)} = \sum_{i=1}^N \sum_{a=1}^d \sum_{\mu=1}^{Nz/2}\left(\pdiff{h_\mu}{x_{ia}}\right)^2
 = Nz,\label{113644_13Feb21}
\end{align}
where we used 
\begin{align}
& \pdiff{h_{ij}}{x_{ka}} = \left(\delta_{ik}-\delta_{jk}\right)
 \frac{x_{ia}-x_{ja}}{\abs{\br_i-\br_j}},
& \sum_{k=1}^N \sum_{a=1}^d\left(\pdiff{h_{ij}}{x_{ka}}\right)^2 = 2.
\end{align}
The RHS is 
\begin{align}
 a \tr \WW = \frac{2a}{Nz}\sum_{i=1}^N\sum_{a=1}^d \sum_{\mu=1}^{Nz/2}\left(\xi_{ia}^\mu\right)^2 = a Nd.\label{113652_13Feb21}
\end{align}
By using Eqs.~(\ref{113356_13Feb21}), (\ref{113644_13Feb21}) and
(\ref{113652_13Feb21}), we get
\begin{align}
 a = \frac{z}{d}.
\end{align}

Next we consider the full matrix including the term proportional to $e$.
Assuming that $\tr\HH=\tr\HH_{\rm MF}$, we get
\begin{align}
 &\tr \left(\HH^{(1)}+\HH^{(2)}\right)  = \tr\left(a\WW + be\II\right)\new
 & \to \tr\HH^{(2)}  = be \tr\II.\label{120149_13Feb21}
\end{align}
The LHS can be calculated as 
\begin{align}
 \tr\HH^{(2)} &= \sum_{i=1}^N\sum_{a=1}^d \sum_{\mu=1}^{Nz/2}h_\mu\pdiff{^2 h_\mu}{x_{ia}^2}\new
 &= 2(d-1)\sum_{\mu=1}^{Nz/2}\frac{h_\mu}{r_\mu}\new
 &= -Nze,\label{120213_13Feb21}
\end{align}
where we used
\begin{align}
& \pdiff{^2 h_{ij}}{x_{ka}^2} =
\left(\delta_{ik} +
\delta_{jk}\right)\frac{r_{ij}^2-(x_{ia}-x_{jb})^2}{r_{ij}^3},\new
&\sum_{k=1}^{N}\sum_{a=1}^d\pdiff{^2 h_{ij}}{x_{ka}^2} =2 \frac{d-1}{r_{ij}}.
\end{align}
From Eqs.~(\ref{120149_13Feb21}), (\ref{120213_13Feb21}) and $\tr\II = Nd$, we get
\begin{align}
b = -\frac{z}{d}.
\end{align}

In summary, we get 
\begin{align}
 \HH_ {\rm MF} = 
 \frac{z}{d}\WW
 -\frac{z}{d}e\II.\label{120053_30Oct20}
\end{align}

\subsection{Eigenvalue distribution}

It is well-known that the eigenvalue distribution of the Wishart matrix
$\WW$ follows the Marchencko-Pastur law~\cite{livan2018introduction}
\begin{align}
 \rho_{\rm MP}(\lambda) = 
 \frac{z}{2d}\frac{\sqrt{(\lambda_+-\lambda)(\lambda-\lambda_-)}}{2\pi\lambda},\
 \lambda_{\pm} = \left(1\pm \sqrt{\frac{2d}{z}}\right)^2.
\end{align}
Let $\be_n$ be an eigenvector of $\WW$, and $\lambda_n^{\rm MP}$ be the
corresponding eigenvalue. Then, we have
\begin{align}
\HH_{\rm MF} \cdot\be_n  = \left(\frac{z}{d}\lambda_n^{\rm MP}-\frac{z}{d}e\right)\be_n,
\end{align}
meaning that $\be_n$ is also an eigenvector of $\HH_{\rm MF}$ and the
corresponding eigenvalue is $\lambda_n = \frac{z}{d}\lambda_n^{\rm
MP}-\frac{z}{d}e$. Therefore, the eigenvalue distribution of $\HH_{\rm
MF}$ is calculated as
\begin{align}
 \rho(\lambda)  = \rho_{\rm MP}(\lambda_{\rm MP})\diff{\lambda_{\rm MP}}{\lambda}
 = \frac{d}{z}\rho_{\rm MP}(d\lambda/z + e).\label{220413_13Sep20}
\end{align}
In particular, the minimal eigenvalue is 
\begin{align}
 \lambda_{\rm min} =
\frac{z}{d}\left(1-\sqrt{\frac{2d}{z}}\right)^2-\frac{z}{d}e.\label{162217_27Oct20}
\end{align}

\subsection{Marginal stability and contact number}
The replica calculation in the limit $d\to\infty$ predicts that the
system is marginally stable near the jamming transition point,
$\lambda_{\rm min}=0$~\footnote{ Here we take the limit $d\to\infty$ so
that $z/d$ remains finite. This is consistent with the numerical result
the jamming transition point where $z\approx 2d$~\cite{ohern2003}.}
~\cite{franz2015universal,kurchan2013exact,parisi2020theory}. By using
the marginal stability and Eq.~(\ref{162217_27Oct20}), we get
\begin{align}
\frac{z(e)}{2d} = \frac{1}{(1-e^{1/2})^2}.\label{220422_13Sep20}
\end{align}
Here we assumed that $z/d$ takes a finite value in the limit
$d\to\infty$, because the numerical results suggest that $z\to 2d$ at
the jamming transition point~\cite{ohern2003}. For $e\ll 1$, we
reproduce the well-known scaling observed by numerical
simulations~\cite{ohern2003}
\begin{align}
z/2d-1\sim 2 e^{1/2}\sim p^{1/2}.
\end{align}
The critical exponent $1/2$ was previously derived by using the
variational argument~\cite{wyart2005}, effective medium
theory~\cite{degiuli2014effects}, and replica
theory~\cite{franz2017universality}, but our result
Eq.~(\ref{220422_13Sep20}) also allows us to access the pre-factor and
non-linear terms. Somewhat surprisingly, Eq.~(\ref{220422_13Sep20})
suggests that $z(e)$ depends only on $e$, and does not depend on the
preparation protocols. It is interesting future work to see if this
property survives in finite $d$.

\begin{figure}[t]
\begin{center}
 \includegraphics[width=9cm]{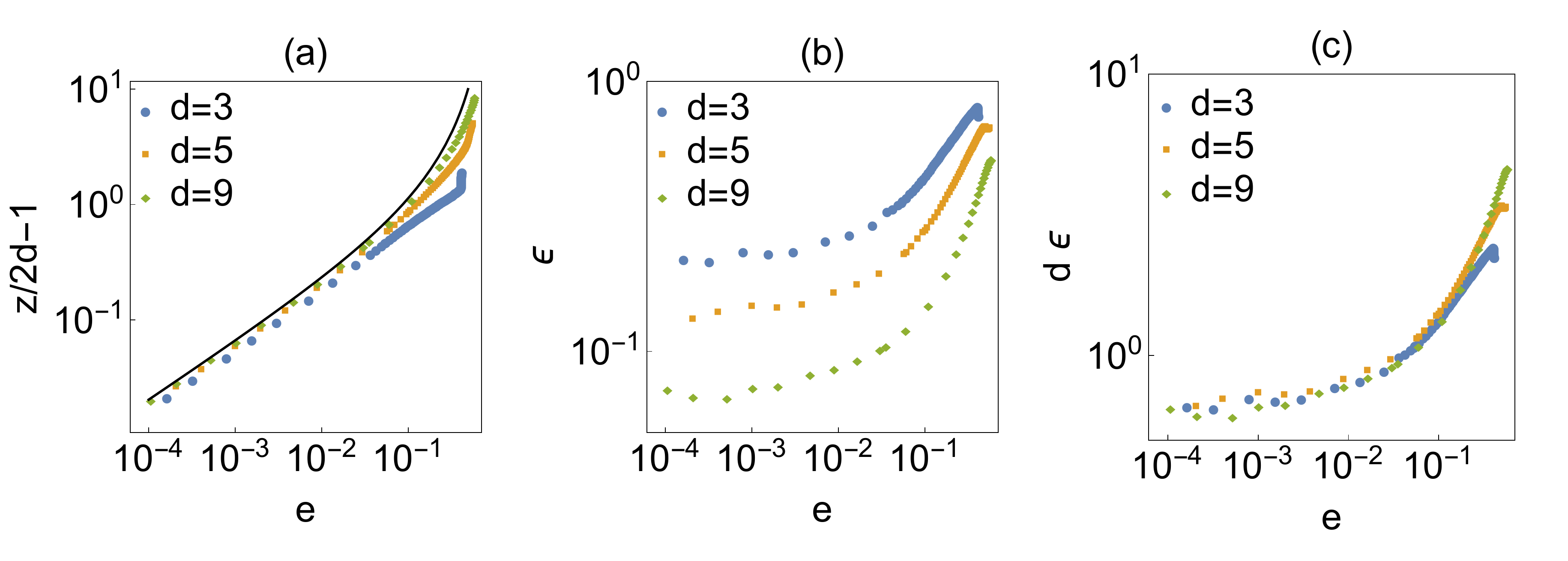} 
\vspace{-0.8cm}
 \caption{(a) $e$ dependence of
 $z$. Markers denote numerical results taken from
 Ref.~\cite{shimada2020low}, while the solid line denotes the
 theoretical prediction. (b) $\epsilon$ for the same data. (c)
 $d\epsilon$ for the same data.}  \label{220525_13Sep20}
\end{center}
\end{figure}

In Fig.~\ref{220525_13Sep20}~(a), we compare Eq.~(\ref{220422_13Sep20})
with numerical results in several spatial dimensions $d$ obtained by
rapid quench from high temperature random configurations. See
Ref.~\cite{shimada2020low} for the details of the numerical simulations.
The theory well agrees with the numerical results for small $e$. For
more quantitative discussion, in Fig.~\ref{220525_13Sep20}~(b), we show
the difference between the results of the theory $z_{\rm the}$ and
simulation $z_{\rm sim}$:
\begin{align}
 \epsilon = \frac{z_{\rm the}/2d-1 - \left(z_{\rm sim}/2d-1\right)}{z_{\rm the}/2d-1}
 = \frac{z_{\rm the}-z_{\rm sim}}{z_{\rm
the}-2d}.
\end{align}
The data collapse onto a single curve if we rescale the vertical axis by
$d$ (Fig.~\ref{220525_13Sep20}~(c)), meaning that the deviation scales
as $\epsilon\sim 1/d$.

\subsection{Vibrational density of states}

\begin{figure}[t]
\begin{center}
 \includegraphics[width=9cm]{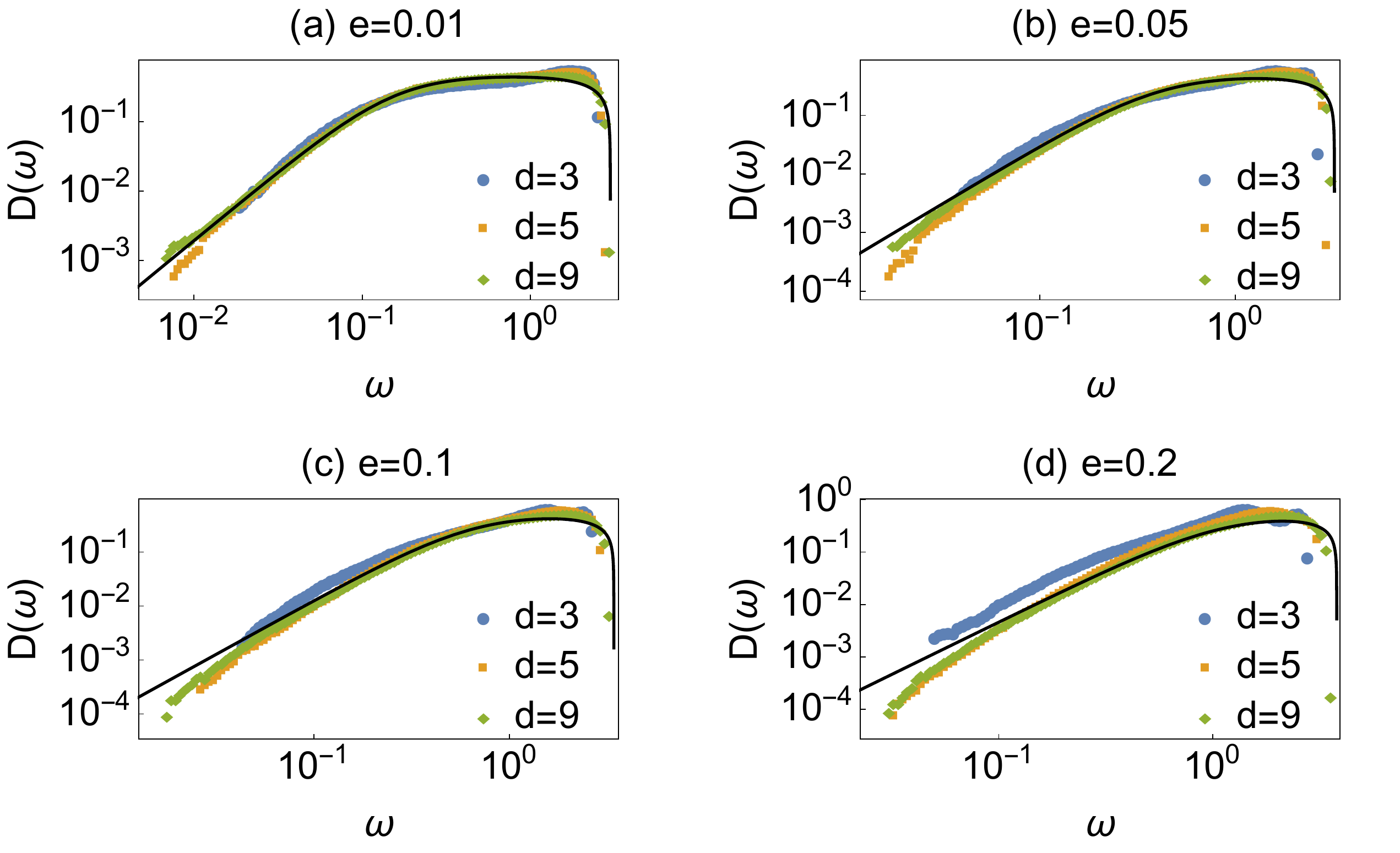} 
\vspace{-0.8cm}
\caption{Density of states
 $D(\omega)$. Markers denote numerical results taken from
 Ref.~\cite{shimada2020low}. The solid lines denote theoretical
 predictions.}  \label{000725_14Sep20}
\end{center}
\end{figure}

By using Eq.~(\ref{220413_13Sep20}), the vibrational density of states
$D(\omega)$ is calculated as ${D(\omega)=2\omega\rho(\lambda=\omega^2)}$.
Although $D(\omega)$ depends on both $z$ and $e$,
Eq.~(\ref{220422_13Sep20}) allows us to eliminate the dependency on
$z$. After some manipulations, we get
\begin{align}
D(\omega) = 
 \frac{\omega^2\sqrt{(1-e^{1/2})^3\left\{8-(1-e^{1/2})\omega^2\right\}}}{2\pi\left\{2e+(1-e^{1/2})^2\omega^2\right\}}.
 \label{133605_17Sep20}
\end{align}
In Fig.~\ref{000725_14Sep20}, we compare the theoretical prediction
Eq.~(\ref{133605_17Sep20}) and numerical results. The results are
consistent near jamming $e=0.01$ even in $d=3$, while there is a visible
deviation for small $\omega$ far from jamming $e=0.2$ even in $d=9$. It
is an interesting future work to see if a better agreement is obtained
in higher $d$.

For $e\ll 1$ and $\omega\ll 1$, we get the following scaling:
\begin{align}
 D(\omega) \sim
\frac{\omega^2\sqrt{\omega_{\rm max}^2-\omega^2}}{2\pi(\omega^2+\omega_*^2)}
 \sim 
 \begin{cases} 
  {\rm const} & \omega\gg \omega_*\\
  \delta z^{-2}\omega^2 & \omega\ll \omega_*,
 \end{cases}
\end{align}
where $\omega_{\rm max}=2\sqrt{2}$, and $\omega_* = \sqrt{2e}\propto
z/2d-1$. 
In particular, $D(\omega)={\rm const}$ at the jamming transition point $e=0$.
The similar results have been previously derived by applying
the effective medium theory to the disordered
lattices~\cite{degiuli2014effects}, and the replica method to the
mean-field models~\cite{franz2015universal,ikeda2020}.

\section{Summary}
\label{170931_26Oct20}

In this work, we calculated the contact number $z$ and vibrational
density of states $D(\omega)$ for harmonic spheres in the large spatial
dimensions $d\to\infty$. Our theoretical results well agree with the
results of the previous numerical simulation in large $d$.  



Our theoretical results are relied on the Ansatz
Eq.~(\ref{131641_25Oct20}), that is, the Hessian of harmonic spheres has
the form of the shifted Wishart matrix. The consistency between our
theoretical results and previous numerical results suggests that the
Ansatz becomes exact in the limit $d\to\infty$.  This result motivates
us to develop more rigorous calculation in $d\to\infty$ without using
the Ansatz, as done in the previous exact calculations for hard
spheres~\cite{kurchan2012exact,kurchan2013exact,charbonneau2014exact}.
We left it as future work.

\acknowledgments

We thank F.~Zamponi
for useful comments. This project has received
JSPS KAKENHI Grant Numbers 21K20355
and 19J20036.

\appendix

\bibliography{reference}

\end{document}